\documentclass[twocolumn,showpacs,preprintnumbers,amsmath,amssymb]{revtex4}

\usepackage{graphicx}
\usepackage{epsfig}
\usepackage{dcolumn}
\usepackage{bm}

\begin{document}

\title{Spin current generation by helical states in a quasi-one-dimensional system.}

\author{Manuel Val\'{\i}n-Rodr\'{\i}guez}
\affiliation{%
Conselleria d'Educaci\'o i Cultura, Govern de les Illes Balears.
07004 Palma de Mallorca, Spain
}%

\date{September 22, 2011}

\begin{abstract}

Time-reversal symmetry and rotational invariance in spin space characterize usual non-magnetic conductors. These symmetries
give rise, at least, to four-fold degenerate multiplets which, by definition, exhibit a null total spin-momentum helicity. Thus, preventing
a net spin transport. A proper choice of geometry along with the intrinsic symmetry of the Bychkov-Rashba spin-orbit interaction
can be exploited to effectively reduce these two spin-related symmetries to the time-reversal one. It is shown that, in an ideal geometry,
a quantum dot with contacts having a specific geometry exhibit a single pair of helical propagating states which makes this system ideal
for pure spin current generation. The strong quantization of the quantum dot's level structure would make this mechanism robust against
temperature effects.

\end{abstract}


\maketitle

In the last decades, the fundamental magnitude used for transport and processing of information has been the electric charge. 
However, the interactions responsible for the charge manipulation require a relatively
high energy consumption and impose restrictive limitations to the operability of
the devices.

In recent years, there has emerged a new branch of condensed matter physics which 
seeks to exploit the internal degree of freedom of the electron rather than 
its charge for information processing, also known as spintronics \cite{zutic,fab,wolf}. The technological advantages of 
using the spin as the information carrier is undoubted since the subtle interactions
involved would allow devices with lower power consumption and better performance.

A major constraint for potential technological applications in semiconductor materials
is that the manipulation of spin must be done through electrical means. Therefore, the 
most feasible strategy for spin manipulation would be to use intrinsic spin-dependent 
interactions that do not require external elements of magnetic nature such as applied
magnetic fields or magnetic impurities, which are difficult to integrate. In this sense,
the spin-orbit interaction is a natural mechanism for the spin manipulation as it is
intrinsically present in a wide variety of semiconductor alloys.

A key concept in the field of spintronics is the pure spin current, i.e., a net flow 
of spin without any net flow of electric charge. A precise control of this magnitude
would allow the processing of information without the heavy operational costs 
associated with the charge drift.

A paradigmatic mechanism for the generation of spin currents based on the spin-orbit
interaction is the intrinsic spin Hall effect \cite{murakami,sinova}. The effect of spin-orbit interaction induces a 
spin-dependent deflection transversal to the motion of the carriers. This generates 
a spin current perpendicular to the charge current defined by the orientation of the momentum.
Although the intrinsic spin Hall effect actually generates a pure spin current, the fact that it is
always accompanied by a charge current limits its applicability in a scheme based
entirely on the spin. Despite this, it provides a method for electrical detection of a 
spin current through its reciprocal effect, the inverse spin Hall effect, since the spin-orbit
interaction induces a transversal charge current when acting on a pure spin current. 

More recently, the discovery of topological insulators \cite{berne,berne2,kane} offers novel 
possibilities for the technological use of the spin. These materials are characterized 
by an inverted band structure due to the high intensity of the spin-orbit interaction. 
In this new state of matter the material behave as insulator on the substrate but 
show conductivity at the edges through localized states. At each edge, the system exists
an odd number of time-reversal conjugated doublets (Kramers doublets) \cite{wu,qi} known in this context 
as helical states because of the correlation between the direction of propagation and the 
spin orientation, i.e., a left-mover with spin up has the same helicity as a right-mover
with spin down. The helicity is opposite at each edge of the system. The oddness of 
time-reversal doublets at each end is of great relevance in the transport properties 
since the scattering through non-magnetic impurities is forbidden leading to 
dissipationless transport \cite{berne2,kane}.

Note also that the odd-helical doublets can  generate pure spin currents since 
each doublet corresponds to states with opposite momentum and spin orientation, thus 
a doublet of helical states cannot produce a net charge transport since the two states have
opposite sense of motion, but a net transport of spin is carried out due to the opposite
spin orientation of the travelling waves. A system with an even number of helical 
doublets will not show this effect since the spin transport properties of a doublet cancel
out with those of the doublet having opposite helicity. Therefore, a system characterized 
by an uncompensated spin-momentum helicity is likely to generate a
pure spin current.

It is the aim of this letter to show that a quasi-one-dimensional system can be engineered 
to exhibit propagating states characterized by an uncompensated spin-momentum helicity and, hence, 
capable of spin current generation. 

Consider a generic conductor without any magnetic or spin-dependent interactions, thus 
preserving the time-reversal symmetry and the rotational invariance in spin space. Regardless of its geometry or dimensionality, this 
system displays, at least, a fourfold degeneracy, except for motionless states defined as those
where the inversion of momentum doesn't change the orbital state. Therefore, such a  system
cannot generate spin currents since the net spin-momentum helicity is constrained to be null by symmetry considerations.
The key point to avoid this restriction is to find a magnitude that differentiates the Kramers doublets.
The degree of freedom introduced by this magnitude would allow to control the transport properties 
of the different doublets.

In a proper geometry, the axial total angular momentum ($J_z = L_z + S_z$) is a quantity that differentiates the
Kramers doublets so it will be the basis for the mechanism of spin current generation. 
The other ingredient needed to take advantage of the doublets' differentiation is a filtering mechanism
that allows the transport through doublets characterized by a particular value of $J_z$ while for other
values ​​the transport is forbidden. In our scheme, the total angular momentum filtering element 
consists of a quantum dot.

\begin{figure}
\includegraphics[width=0.45\textwidth]{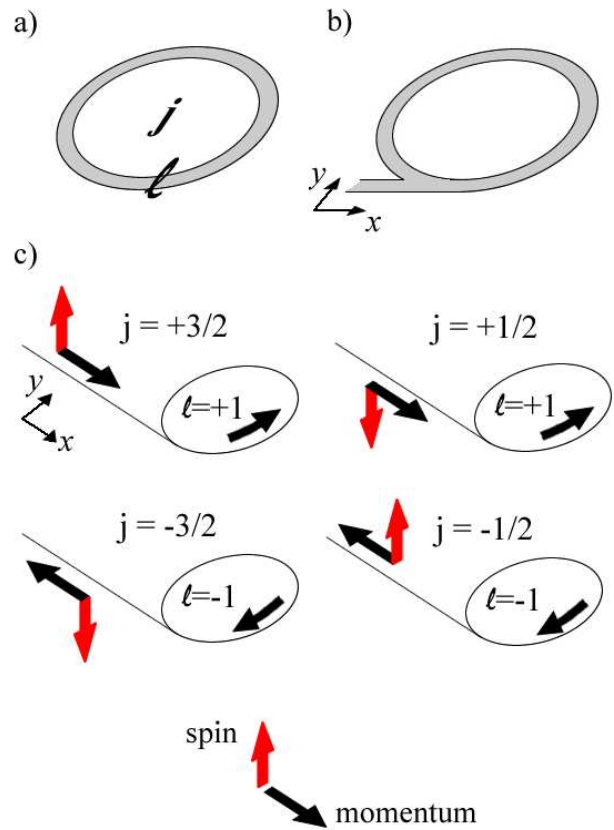}
\caption{a) Schematic representation of a non-magnetic ring contact ($\ell$ - orbital angular momentum) surrounding a quantum dot with
spin-orbit interaction ($j$ - total angular momentum). b) Schematic representation of an asymmetric wire-ring contact. c) Four-fold degenerate
multiplet in the ideal one-dimensional wire-ring contact.
}
\label{rest}
\end{figure}

A quantum dot exhibits a full quantization of its level structure and the states of the different energy subbands are characterized
by different orbital properties which can be exploited to control the transport through them. The typical orbital energy spacing 
of the quantized spectrum ranges from a few meVs in lateral quantum dots to several tens of meVs for vertical
 quantum dots \cite{revhanson, koneman}, the latter being of interest for the present work. In particular, 
we will consider the conduction band spectrum
 of a two-dimensional quantum dot with Bychkov-Rashba spin-orbit interaction \cite{rash}. For a wide variety of 
semiconductor materials this can be accurately described by an effective mass Hamiltonian

\begin{equation}
{\cal H} = \frac{p_x^2+p_y^2}{2m^*}+V(x,y)+\frac{\alpha}{\hbar}\left(p_y\sigma_x-p_x\sigma_y\right)
\end{equation}

where $m^*$ represents the conduction band effective mass, V(x,y) is the confining potential in the plane 
and $\alpha$ is the intensity of the spin-orbit interaction which strongly varies for the different materials.
A general property of this Hamiltonian is that, when the potential has circular symmetry, the axial total 
angular momentum is preserved $\left[{\cal H},J_z\right]=0$. Since the spin-orbit coupling doesn't break
the time-reversal symmetry, the quantum dot also shows the Kramers degeneracy. In particular, the
Kramers doublets are characterized by an opposite value of the total angular momentum eigenvalue.
Another common feature of symmetric quantum dots is that the lowest energy shell is characterized
by states with $j = \pm1/2$. In the well-known case without spin-orbit interaction 
this corresponds to the doublet of states corresponding to the first radial solution ($n = 0$) with zero orbital 
angular momentum ($\ell = 0$) and spin up ($j=+1/2$), down ($j=-1/2$). When the spin-orbit interaction is
taken into account, neither the orbital angular momentum nor the spin are conserved; the only good 
quantum number is the axial total angular momentum.

The effect of spin-orbit interaction on the spectrum of confined systems is small for typical semiconductor
 materials, reaching at most a fraction of meV. Compared with the typical orbital energy scale of a
vertical quantum dot (some tens of meVs) we can safely consider the Bychkov-Rashba spin-orbit
interaction as a perturbation to the Hamiltonian ${\cal H}_0 = (p_x^2+p_y^2)/2m^*+V(x,y)$ \cite{alei}. Apart from 
normalization constants, to first order, the wavefunctions read \cite{lolodef},
 \begin{eqnarray}
&& {\bf \Phi}_{nj\uparrow}({\bf r})\approx \phi_{nj}({r})e^{i(j-1/2)\phi} \left(
\begin{array}{c}
1 \\
-\alpha r e^{i\phi}
\end{array}
\right), 
\nonumber\\
&&{\bf \Phi}_{nj\downarrow}({\bf r})\approx \phi_{nj}({r})e^{i(j+1/2)\phi} \left(
\begin{array}{c}
\alpha r e^{-i\phi} \\
1
\end{array}
\right)
\end{eqnarray}
where the spinorial part is represented in the usual $\sigma_z$-basis. Due to the relative weakness of the
spin-orbit interaction, the spin orientation is mainly aligned along the z-axis of the spin space. A weak 
spin texture with radial orientation is displayed for the in-plane spin distribution, which adiabatically 
follows the orientation of the momentum-dependent effective magnetic field of the Bychkov-Rashba term. 
Despite the spin orientation is not well-defined, we can safely consider these states to be 'up' and 'down', ignoring the subtleties 
of their spin distribution.

Let us consider a ring contact surrounding the quantum dot, as sketched in
Fig. 1 a). This contact is supposed to be non-magnetic and, therefore,
the spin and orbital degrees of freedom are decoupled. Since the contact
Hamiltonian commutes with $L_z$, its eigenstates
have well-defined orbital angular momentum '$\ell$'.
The spin is randomly oriented, so we can choose the $\sigma_z$-basis as usual.

When the system is considered as a whole (contact + quantum dot), the separate conservation of the orbital
angular momentum and a spin  is not ensured in the transitions between the two
subsystems because of the effect of the spin-orbit coupling. However, since the the in-plane circular 
symmetry is maintained, the Hamiltonian of the composed system still preserves the axial total angular
momentum. The constraint imposed by this symmmetry is the key for the proposed mechanism of spin current generation.

\begin{figure}
\includegraphics[width=0.45\textwidth]{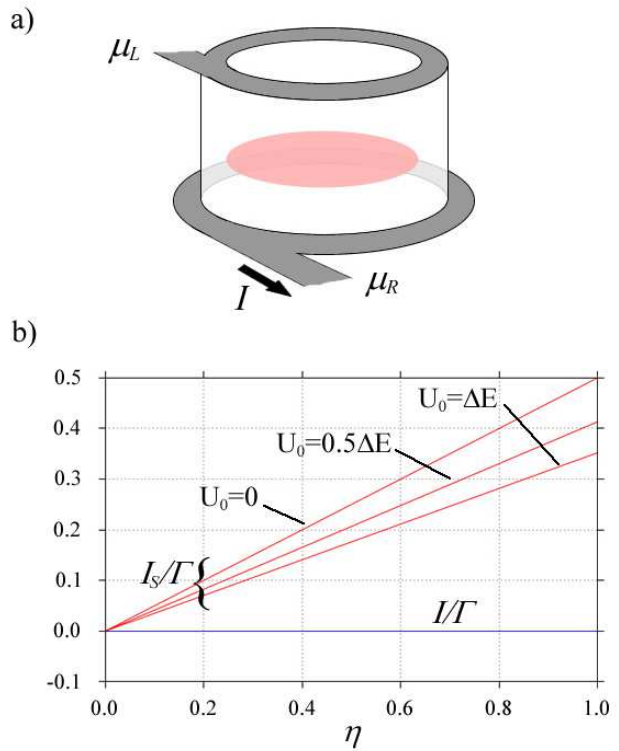}
\caption{a) Three-dimensional representation of the system composed of a quantum dot embedded between two 
wire-ring contacts. b) Charge and spin currents through the dot vs. the efectiveness parameter $\eta$ for different
values of the charging energy $U_0$ ($e=\hbar=1$). A value for the temperature $k_BT=0.3\Delta E$ has been used.
}
\label{rest}
\end{figure}

To exploit this conservation law is necessary that the eigenstates in the contact are differentiated by their total angular momentum. 
This can be achieved by attaching a wire tangentially to the ring as shown in Fig. 1 b). Using such an asymmetrized
contact geometry makes incoming and outgoing states to have different '$j$' in the ring depending on their spin orientation.
To get a clear picture, let's consider this composed contact in the analytically solvable one-dimensional limit. This approximation is physically meaningful, 
if the contact is narrow enough so that the transversal motion in the contact is frozen to the first subband for the energy range of
interest. Using this approximation, the contact Hamiltonian is straightforward to solve (see a schematic representation in Fig.1c).
An incoming electron in the wire, $e^{ikx}$, corresponds to an orbital angular momentum eigenfunction
in the ring (of radius $R$) with integer and positive eigenvalue $\ell$, if the condition $k=\ell/R$ is fulfilled.
On contrary, an outgoing electron, $e^{-ikx}$, corresponds to an orbital angular momentum eigenfunction
in the ring with a negative eigenvalue, if the same condition $k=\ell/R$ is fulfilled. In particular, for $k=0$ the eigenfunction correponds to that given
by $\ell=0$ in the ring, leading to a global constant wavefunction. For all other values
of $k$ (which don't fulfill the condition $k=\ell/R$) there is a total reflection of the wavefunction
at the wire-ring interface.

Note that, if the transport through the quantum dot is restricted to the first energy shell (characterized by  $\mid j\mid = 1/2$),
the only states in the contact that can tunnel to the quantum dot are those having $j =+1/2$ or $j=-1/2$
because of the total angular momentum conservation. In Fig. 1 c), there are represented the four-fold degenerated states in
the contact compatible with $\mid j\mid = 1/2$, which correspond to $\ell=\pm 1$ with spin 'up' and 'down'. Although degenerated
in energy they are distinguished by $j$: the four states are organized in two doublets of Kramers conjugated states, one characterized 
by $\mid j\mid = 1/2$ while the other by $\mid j\mid = 3/2$. Consequently, the tunneling from the $\mid j\mid = 3/2$ states to the quantum 
dot is forbidden. The $\ell=0$ states, having lower energy, are also characterized by $\mid j\mid = 1/2$, however they will be ignored since 
they constitute global constant wavefunctions that don't carry any charge or spin.

In the geometry represented in Fig. 2 a), the quantum dot operated at the lowest energy shell acts as a total 
angular momentum filter, allowing the propagation of only one pair of helical states between the left and right 
contacts. In this way, the symmetry of the whole system is effectively reduced to the time-reversal symmetry 
through this total angular momentum blocking mechanism.

Because of the net spin-momentum helicity characterizing the propagating states, this system can naturally 
generate pure spin currents. It can be checked by including this effect in a simple transport model. 
Let's suppose that the orbital subbands of the dot are strongly spaced so we can restrict to the transport through
the lowest one. In this context, four multi-electron states have to be considered: no electron
occupancy of the dot $(0,0)$; one electron occupying the $j=+1/2$ state $(1,0)$; one electron
occupying the $j=-1/2$ state $(0,1)$; and full subband occupation $(1,1)$. In the latter case,
the Coulomb interaction is taken into account through a constant charging energy $U_0$. The master
equation for the occupancy probabilities in the dot is solved \cite{dattabook} using the Fermi golden
rule for the transition rates between the different multielectron states i,j 
\begin{equation}
\Gamma_{ij}=\sum\limits_{k=L,R} \Gamma_k[f_{ijk}\delta_{n_i,n_j+1}+(1-f_{jik})\delta_{n_i,n_j-1}]
\end{equation}
where $\Gamma_{L,R} =2\pi D_{L,R} |\gamma_{L,R}|^2$ and $f_{ijk}=1/(\exp[(E_i-E_j-\mu_k)/k_BT]+1)$.
We keep the density of states $D_{L,R}$ as energy-independent for both the left and right contacts,
while the tunneling couplings $\gamma_{L,R}$ depend strongly on the value of the total angular 
momentum of the contact states. For completeness of the model, it is introduced a
parameter $\eta$ which quantifies the mechanism efficiency by including a prefactor $1-\eta$ in 
the tunneling rate $\gamma_{L,R}$ for the blocked transitions: when $\eta=0$ there are no transitions
forbidden and no effect of the blocking mechanism is expected, while for $\eta=1$ only those transitions 
allowed by the total angular momentum conservation contribute to the transport.

In Fig. 2 b) it is shown the dependence on the effectiveness parameter $\eta$ of the charge current $I$
($I = I_{\uparrow}+ I_{\downarrow}$) and the spin current $I_s$ ($I_s=I_{\uparrow}- I_{\downarrow}$) for
different values of the charging energy $U_0$ when no bias is applied ($\mu_L=\mu_R=E_0, E_0$ corresponds to the energy of the
lowest orbital subband) and symmetric coupling
is considered ($\Gamma_L=\Gamma_R=\Gamma$). The mean subband energy spacing in the dot $\Delta E$ is used 
as the reference unit for the other energy magnitudes $U_0, k_BT$. The symmetric occupancy probabilities in the dot
results in a linear dependence of the currents in the parameter '$\eta$'. As expected, the charge current is null in all cases
since no difference of electrochemical potential is considered. However, the relevant result is that as the blocking mechanism
is activated ($\eta$ increasing) there is a net flow of spin without a net charge flow. This clearly reflects the fact that only
a single pair of travelling helical states are allowed. 

A major advantage of using a quantum dot as the active element for this total angular momentum blocking  
mechanism is the relatively high level spacing that characterizes its level structure ($\Delta E$). This spacing depends basically
on the dot size, which can be controlled and engineered. For a vertical semiconductor quantum dot the 
orbital level spacing can be as large as a few tens of meV. Since the dot is operated at the lowest energy
subband, a relatively strong thermal smearing would be needed to open new conducting channels, making this 
mechanism robust against temperature effects. A more complete model including transport through
different orbital subbands of the dot, multiple contact transversal subbands and phonon coupling should be used to 
obtain quantitative results at relatively high temperatures. However, phonon coupling is not expected to introduce
significant effects on the spin current since the coupling of the spin degree of freedom is limited locally
to the quantum dot where the Bychkov-Rashba is present. It has been shown that phonon coupling is negligible
in a quantum dot with Bychkov-Rashba interaction at zero magnetic field \cite{stano}.
Another advantadge of this mechanism is that it does not necessarily require a strong spin-orbit interaction
since it relies in the symmetry of the Bychkov-Rashba interaction rather than its intensity. 
Therefore, materials having a weak spin-orbit intensity can also show the effect. Nevertheless, the smaller
the spin-orbit intensity, the smaller spin current generation,
since the relevant transitions involved in the transport depend strongly on the spin-orbit intensity.

In summary, it has been shown in an ideal geometry that a proper choice of geometry in conjunction with the
intrinsic symmetry of the Bychkov-Rashba spin-orbit interaction can be exploited to engineer a quantum dot  based system that exhibits
a single pair of time-reversal propagating states. The net spin-momentum helicity characterizing this system would
allow for pure spin current generation. It is expected this mechanism to be robust against temperature effects due to the 
strong quantization of the quantum dot's level structure.

\end{document}